\documentclass[traditabstract]{aa} 
\usepackage{graphicx}
\usepackage{natbib}
\newcommand{\Mjup}{\ensuremath{\,{M}_{\rm Jup}}}              
\newcommand{\Rjup}{\ensuremath{\,{R}_{\rm Jup}}}              
\newcommand{\pjup}{\ensuremath{\,\rho_{\rm Jup}}}                 
\newcommand{\Porb}{\ensuremath{P_{\rm orb}}}                      
\newcommand{\mss}{\,m\,s$^{-2}$}                                  
\newcommand{\Vsys}{\ensuremath{V_\gamma}}                         

\usepackage{amsmath,amssymb,graphicx}
\begin{document}

\title{Kepler-432\,b: a massive planet in a \\
highly eccentric orbit transiting a red giant}

\titlerunning{Kepler-432\,b: a planet transiting a red giant}

   \author{
          S. Ciceri\inst{1}
          \and
          J. Lillo-Box\inst{2}
          \and
          J. Southworth\inst{3}
          \and
          L. Mancini \inst{1}
          \and
          Th. Henning\inst{1}
          \and
          D. Barrado\inst{3}}
          {
\institute{
Max Planck Institute for Astronomy, K\"{o}nigstuhl 17, D-69117, Heidelberg, Germany \\
    \email{ciceri@mpia.de}
         \and
Departamento de Astrof\'{i}sica, Centro de Astrobiolog\'{i}a
(CSIC-INTA), 28691 Villanueva de la Ca\~{n}ada (Madrid), Spain
                 \and
Astrophysics Group, Keele University, Staffordshire, ST5 5BG, UK
    }

  \abstract
{We report the first disclosure of the planetary nature of
Kepler-432\,b (aka \emph{Kepler} object of interest KOI-1299.01).
We accurately constrained its mass and eccentricity by
high-precision radial velocity measurements obtained with the CAFE
spectrograph at the CAHA 2.2-m telescope. By simultaneously
fitting these new data and \emph{Kepler} photometry, we found that
Kepler-432\,b is a dense transiting exoplanet with a mass of
$M_{\mathrm{p}} = 4.87 \pm 0.48$\Mjup\ and radius of
$R_{\mathrm{p}} = 1.120 \pm 0.036$\Rjup. The planet revolves every
$52.5$\,d   around a K giant star that ascends the red giant
branch, and it moves on a highly eccentric orbit with $e = 0.535
\pm 0.030$. By analysing two NIR high-resolution images, we found
that a star is located at $1.1^{\prime\prime}$ from Kepler-432,
but it is too faint to cause significant effects on the transit
depth. Together with Kepler-56 and Kepler-91, Kepler-432 occupies
an almost-desert region of parameter space, which is important for
constraining the evolutionary processes of planetary systems.}

\keywords{stars: planetary systems -- stars: fundamental parameters -- stars: individual: Kepler-432 --  techniques: spectroscopy}

\maketitle

\section{Introduction}
\label{sec_1}

Since its first data release \citep{borucki2011a}, the
\emph{Kepler} spacecraft has been the most productive
planet-hunting mission. It has allowed the discovery of
over 4\,000 exoplanet candidates to date, with a very low false positive
frequency at least for small planets (e.g. \citealp{marcy2014,
fabrycky2014}).  The false-positive rate is higher
($\sim70\%$) for \emph{Kepler}'s giant stars \citep{sliski2014}.

One of the best ways to unequivocally prove the planetary nature
of a transiting object is to obtain radial velocity (RV)
measurements of the parent star, which also allows precise
constraints on the mass of the planet. Unfortunately, the host
stars of most of the \emph{Kepler} candidates are too faint or
their RV variation is too small to determine the mass of the planets
with current spectroscopic facilities. Nevertheless, considerable effort is made to observationally characterize
many interesting \emph{Kepler} candidates (e.g.
\citealp{hebrard2013,howard2013,pepe2013}) and develop new
instruments with higher resolution and better performance (see
\citealp{pepe2014} for a comprehensive review).

Thanks to the extremely high photometric precision of the
\emph{Kepler} telescope, other methods such as \emph{\textup{transit-timing variation}} (e.g.\ \citealp{holman2010, steffen2013,
xie2013}) and \emph{\textup{orbital brightness modulation}}
(e.g. \citealp{charpinet2012,quintana2013,faigler2013}) have been
adopted to confirm the planetary nature of candidate objects.
Using the latter method, \citet{huber2013a} detected two planets
in the Kepler-56 system, while \citet{lillobox2014a} confirmed the
hot-Jupiter Kepler-91\,b, whose planetary nature was also recently
supported by an independent study based on multi-epoch
high-resolution spectroscopy \citep{lillobox2014c}. Kepler-56\,b,c
and Kepler-91\,b were found to be the first transiting planets
orbiting giant stars.

Up to now, more than 50 exoplanets have been detected around
evolved giants with Doppler spectroscopy, and their general
characteristics are different from those found
orbiting main sequence (MS) stars. According to the study of
\citet{jones2014a}, they are more massive, prefer low-eccentricity
orbits, and have orbital semi-major axes of more than $0.5$\,au
with an overabundance of between $0.5$ and $0.9$\,au. Furthermore,
the correlation between \emph{\textup{stellar metallicity}} and
the \textup{\textup{\emph{\textup{number of planets}} }}seems to be reversed compared with MS
stars, even though there is still an open debate on this matter
(see discussion in  \citealp{jones2014a}). In this context, the
discovery of more exoplanets around evolved stars is vital to
enlarge the sample and better characterize the statistical
properties of these planetary systems. The cases in which the
parent stars are K or G giants, which are known to evolve from F-
and A-type MS stars, are also very interesting for planet
formation and evolution theories and help to form a better demographic
picture of planets around early-type stars.

Here we describe the confirmation via RV measurements of the
transiting planet Kepler-432\,b (aka KOI-1299.01), which we show
to be a massive gas giant moving on a very eccentric orbit around
an evolved K giant that is ascending the red giant branch. Both
Kepler-432\,b and Kepler-91\,b are on tight orbits and present
physical characteristics that deviate from the systems detected so
far by the RV method.

\section{Observations and data analysis}
\label{sec_2}

Kepler-432 was continuously monitored by \emph{Kepler} from May
2009 to May 2013, being observed in all the 17 quarters in
long-cadence mode and during 8 quarters in short-cadence mode. It
was recognized as a \emph{Kepler} object of interest by
\citet{borucki2011b} after it showed a periodic dimming in the
light curve every $52.5$\,d. A subsequent study of \emph{Kepler}
candidates by \citet{huber2013}, making use of the
asteroseismology technique, refined some parameters of this
system. These were updated by \citet{burke2014}. We summarise
relevant parameters in Table \ref{phot-par}.

\begin{table}
\centering %
\caption{Photometric and physical properties of the host star and transit signal from previous studies.} %
\tiny %
\begin{tabular}{ccc}
\hline %
\hline %
Parameter & Value & Reference \\
\hline %
$R$ (mag)                     & 12.135                 & NASA Archive         \\
$K_{p}$ (mag)                 & 12.183                 & NASA Archive         \\
Depth (ppm)                   & 914                    & NASA Archive         \\
Duration (hr)                 & 14.7951                & NASA Archive         \\
$R_\star$ ($R_{\sun}$)        & $4.160 \pm 0.120$      & \cite{huber2013} \\
$M_\star$ ($M_{\sun}$)        & $1.353 \pm 0.101$      & \cite{huber2013} \\
$\rho_\star$ (g\,cm$^{-3}$)   & $0.02650 \pm 0.00049$  & \cite{huber2013} \\
\hline
\end{tabular}
\tablefoot{The data taken from the NASA Exoplanet Archive are available on \texttt{http://exoplanetarchive.ipac.caltech.edu}.}
\label{phot-par}
\end{table}

\subsection{CAFE data}
\label{sec_2.1}

In 2012 we started a programme to confirm a subset of \emph{Kepler}
candidates via spectroscopic follow-up observations. For this
purpose, we used the Calar Alto Fiber Echelle (CAFE)
spectrograph mounted on the 2.2 m Calar Alto telescope. CAFE is an
echelle spectrograph capable of achieving an average resolution of
$R=63\,000 \pm 4\,000$ in the optical regime. The nominal
precision in measuring RVs of stellar objects, tested on known
exoplanet host stars (\citealp{aceituno2013, lillobox2014a}), is a few tens of m\,s$^{-1}$, sufficient to detect a
signal caused by a close-in Jupiter-like planet.

We obtained 28 spectra of Kepler-432 during 16 nights of
observations carried out in the 2013 and 2014 seasons, during the
best visibility time of the \emph{Kepler} field. The exposure time
was 1800\,s for most spectra, but increased to 2700\,s for six
spectra to compensate for the presence of thin clouds and veils.
Each spectrum was extracted from the raw data using the pipeline
provided by the Calar Alto observatory based on the extant R3D
pipeline developed by \citet{sanchez2006}. In brief, each order of
a spectrum is extracted from the flat-fielded and debiased science
image thanks to a continuum image that traces the orders along the
pixels. Each spectrum is calibrated in wavelength using the lines
of a ThAr spectrum taken after the science frame.

RV measurements were obtained by cross-correlating each observed
spectrum with a synthetic spectrum created from the stellar
parameters found in the literature. In particular, the
cross-correlation was made order by order, and the final RV
measurement was the median value of all those obtained
(\citealp{mueller2013}). The RV values estimated are listed in
Table\,A.1 together with their relative uncertainties. Finally,
to better characterize Kepler-432, we combined several CAFE spectra
of the star to obtain one with a high signal-to-noise ratio, from which,
following the methodology described in \cite{fossati2010}, we
inferred the effective temperature of the star $T_{\mathrm{eff}}$
and its surface gravity $\log{g}$, see Table\,\ref{tab:absdim}. We
first fitted the RV data by using the package Systemic Console 2
\citep{meschiari2009}. To obtain the uncertainties relative to the
fitted parameters, we performed bootstrapping and MCMC
simulations and adopted the higher values of the uncertainties
found with the two methods. The different orbital solutions
obtained from the bootstrapping simulations show a clear
preference for an eccentric orbit.

As a sanity check, we obtained a Lomb-Scargle periodogram of the
RVs without considering the transit times from the photometry.
Among the first three peaks we found $P=51.95$\,d (with a
false-positive probability of 0.0136), a value similar to that
obtained from the \emph{Kepler} photometry.

%
\begin{figure}%
\centering
\includegraphics[width=9.0cm]{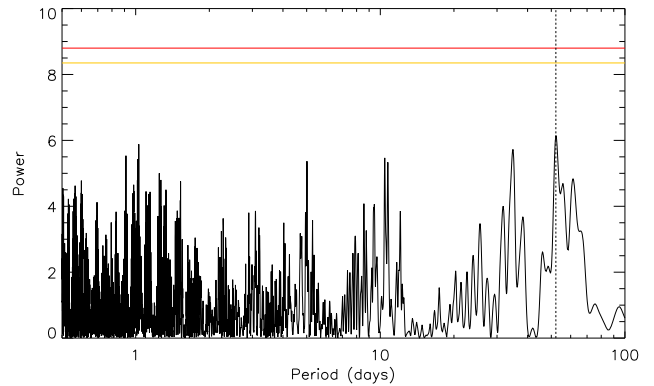}
\caption{Lomb--Scargle periodogram obtained from the CAFE RV
measurements for Kepler-432. The dashed line highlights the orbital
period of Kepler-432\,b. The red and orange lines represent the 5\%
and 10\% false-alarm probability.} \label{period}
\end{figure}

\subsection{Excluding false-positive scenarios}
\label{subsec_2.3}
%
To rule out the possibility that Kepler-432 is a blended stellar
binary system that mimics the observable properties of a
transiting planet system, we analysed two high-resolution images
of Kepler-432 in $J$ and $K$ bands that were obtained with the
NIRC2 imager mounted on the Keck\,II telescope, used in adaptive
optics mode\footnote{The images were published by David Ciardi on
the Community Follow-up Observing Program (CFOP),
https://cfop.ipac.caltech.edu/home/.}. In these images, there is a
clear detection of a nearby star at 1.1\,arcsec, which is much
fainter than Kepler-432\,A, with $\Delta_{J}= 5.68 \pm 0.04$\,mag
and $\Delta_{K}= 5.19 \pm 0.01$\,mag. We translated these
differences into $K_{\mathrm{p}}$-band magnitudes by using the
formulae from \citet{howell2012}, obtaining that the component B
is $6.68 \pm 0.17$\,mag fainter than Kepler-432\,A in the
\emph{Kepler} band. Finally, using the relations from
\citet{lillobox2014b}, we estimated the dilution effect of this
faint star on the depth of the transit events, finding a
correction of 0.01\,$R_{\oplus}$ for the radius of the eclipsing
object. This correction is much smaller than the uncertainty in
our measurement of the radius of Kepler-432\,b. Another
possibility that we have to consider is that the B component might
be an eclipsing binary. However, since we detected an RV signal of
a planetary-mass object with the same periodicity as the transit
signal and the companion is very faint, this scenario is very
unlikely. Instead, the most probable scenario is that the planet
is orbiting component A and that B only acts as a diluting source,
having very weak implications on the planet (and orbital)
properties derived from the light curve.

Another possible source of false positives is stellar activity,
which could mimic the presence of a planetary body in the RV
signal. To rule out this possibility as well, we determined the
bisector velocity span (BVS) from the same spectra from which we
obtained the RV measurements. The BVS values are plotted in
Fig.\,\ref{bisector} together with the best-fitting line, which is
consistent with a horizontal line. This means that we did not find any
significant correlation between RV and BVS.

\begin{figure}
\centering
\includegraphics[width=9.0cm]{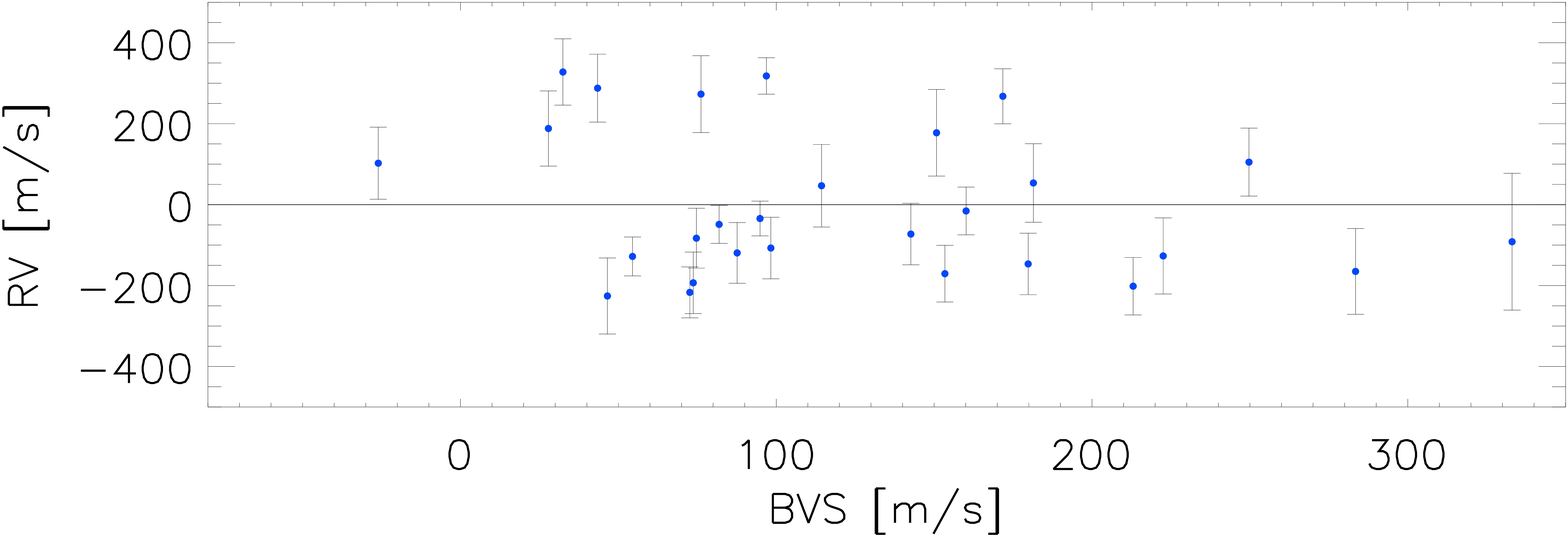}
\caption{Radial velocity (RV) versus bisector velocity span (BVS) for Kepler-432}%
\label{bisector}
\end{figure}

\section{Physical properties of the system}
\label{sec_3}

\begin{figure}
\centering
\includegraphics[width=\columnwidth]{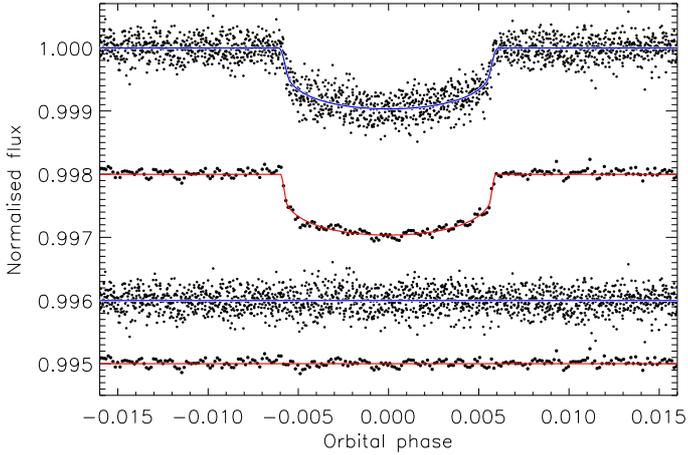}
\caption{\emph{Kepler} long-cadence (top light curve) and short-cadence (bottom light curve) data around transit.
The {\sc jktebop} best fits are shown using solid lines. The residuals of the fits are shown offset towards the base
of the figure. We phase-binned the short-cadence data by a factor of 100 to make this plot clearer.}
\label{fig:lc}
\end{figure}

\begin{figure}
\centering
\includegraphics[width=\columnwidth]{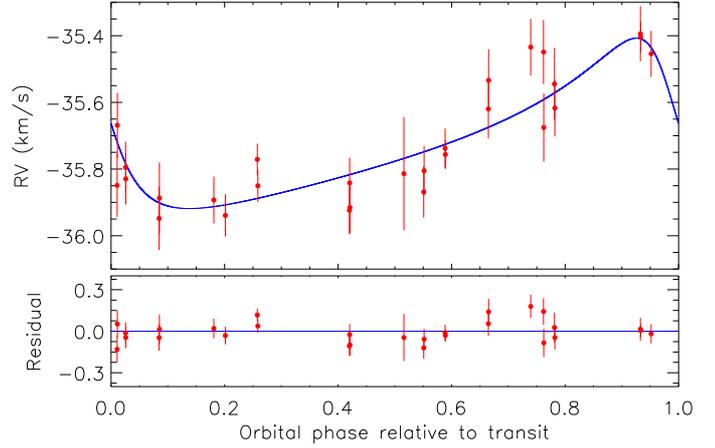}
\caption{Upper panel: phased RVs for Kepler-432 red points) and the best fit from {\sc jktebop} (blue line).
Lower panel: residuals of RVs versus best fit.}
\label{fig:rv}
\end{figure}

To determine the physical parameters of the system, we
simultaneously modelled the \emph{Kepler} photometry and the CAFE
RVs using the {\sc jktebop} code (see \citealt{southworth2013} and
references therein). The parameters of the fit were chosen to be
the fractional radii of the two objects ($r_\star =
\frac{R_\star}{a}$ and $r_{\rm p} = \frac{R_{\rm p}}{a}$ where $a$
is the orbital semi-major axis), orbital inclination $i$, orbital
period \Porb, reference transit midpoint $T_0$, velocity amplitude
$K_\star$, systemic velocity \Vsys\ of the star, the eccentricity
($e$), and argument of periastron ($\omega$) expressed using the
combinations $e\cos\omega$ and $e\sin\omega$.

The \emph{Kepler} long- and short-cadence data were each
converted from flux to magnitude units. Data with more than two transit
durations from a transit midpoint (approximately 95\% of the
data points for both cadences) were rejected to aid computational
efficiency. Each transit was rectified to zero differential
magnitude by subtracting a linear or quadratic polynomial
trend versus time, fitted to the out-of-transit data points. The
short-cadence data were additionally treated by iteratively rejecting
$3\sigma$ outliers, totalling 1.2\% of the data points. Error
bars
for the data for each cadence were assigned to force a reduced
chi-squared of $\chi^2_\nu = 1.0$. The radial velocity error
bars
were scaled by $\sqrt{1.8}$ to achieve the same goal.

The very low ratio of the radii means that the transit is shallow
and the partial phases (ingress and egress) are short. Their
length is poorly determined by the data, leading to $i$ and
$r_\star$ being highly correlated. The solution is indeterminate
without outside constraints. Fortunately, the asteroseismic
density from \citet{huber2013} can be used to rescue the
situation: under the assumption that $M_\star \gg M_{\rm p}$ , the
density is directly related to $r_\star$ \citep{seager2003}. We
therefore fixed $r_\star$ at the value for the known density
(Table\,\ref{phot-par}) and fitted for $i$. Limb darkening was
specified using the quadratic law, whose coefficients were fixed
at the theoretical values given by \citet{sing2010}. We also
assumed that, neglecting the detected B component, no other light
came from the planet or from any additional object along the same
sightline. The low sampling rate of the long-cadence data was
dealt with as in \citet{southworth2011}.

The best fits are shown in Figs.\ \ref{fig:lc} and \ref{fig:rv};
the scatters around the best fits were 0.19 mmag and 0.42 mmag for
the long- and short-cadence. To determine error
estimates, we ran Monte Carlo and residual-permutation
\citep{southworth2008} simulations and adopted the larger of the
two error bars for each parameter. We also needed to account for
the uncertainty in $r_\star$. We did this by calculating solutions
with $r_\star$ fixed at its asteroseismic value $\pm$ its error
bar
to determine the effect on each parameter, and added this in
quadrature to the uncertainty from the Monte Carlo and
residual-permutation simulations.

The result of this process was values for $r_\star$, $r_{\rm p}$,
$i$, \Porb, $T_0$, $K_\star$, \Vsys\ and $e$. Independent results
were calculated for both cadences and found to be consistent. We
adopted those from the short-cadence data because they yield
parameter values with a better precision. The final physical
properties of the system were then calculated using standard
formulae, and the uncertainties were propagated with a Monte Carlo
approach. These results are collected in Table\,\ref{tab:absdim}.

\begin{table} \centering \caption{\label{tab:absdim}
Measured properties of sytem Kepler-432.}
\begin{tabular}{l l}
\hline\hline
Parameter               & Value \\
\hline
$T_{\mathrm{eff}}$ (K)     & 4850  $\pm$ 100            \\
$\log{g}$    (cgs)           & 3.0     $\pm$ 0.5            \\
$T_0$ (BJD/TDB)         & 2455477.02906 $\pm$ 0.0014 \\
$P$ (d)                 & 52.50097 $\pm$ 0.00021 \\
$K_\star$ (km\,s$^{-1}$) & 0.256  $\pm$ 0.021 \\
\Vsys\  (km\,s$^{-1}$)  & -35.73 $\pm$ 0.014 \\
$e\cos\omega$           & 0.256  $\pm$ 0.071 \\
$e\sin\omega$           & 0.469  $\pm$ 0.038 \\
$e$                     & 0.535  $\pm$ 0.030 \\
$\omega$ (degrees)      & 61.3   $\pm$ 7.9 \\
$r_\star$               & 0.06374 $\pm$ 0.00039 \\
$r_{\rm p}$             & 0.001763 $\pm$ 0.000022 \\
$i$ (degrees)           & 88.9 $\pm$ 1.3 \\
\hline
$M_{\rm p}$ (\Mjup)     & 4.87 $\pm$ 0.48 \\
$R_{\rm p}$ (\Rjup)     & 1.120 $\pm$ 0.036 \\
$g_{\rm p}$ (\mss)      & 96 $\pm$ 11 \\
$\rho_{\rm p}$ (\pjup)  & 3.46 $\pm$ 0.48 \\
$a$ (au)                & 0.3034 $\pm$ 0.0089 \\
\hline
\end{tabular} \end{table}

\section{Results and conclusions}
\label{sec_4}

\begin{figure}
\centering
\includegraphics[width=8.4cm]{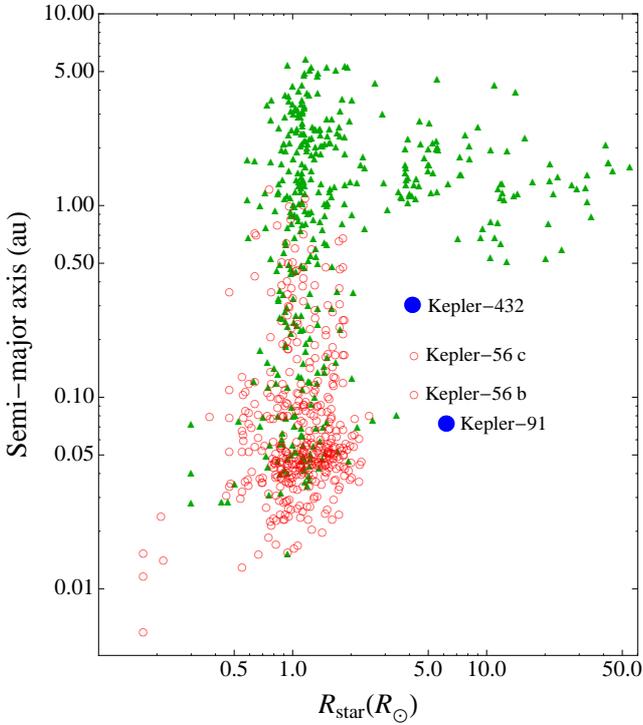}
\caption{Stellar radii and semi-major axis of the orbits of known
planetary systems. Green triangles denote systems found by the RV
method, while red circles are for those found by transit method.
Blue points highlight the positions of Kepler-432 and Kepler-91.}
\label{fig:diagram}
\end{figure}

We confirmed the planetary nature of Kepler-432\,b, a planet with
a mass of $4.87 \pm 0.48\,\Mjup$ and a radius of $1.120 \pm
0.036\,\Rjup$, orbiting a K giant that is ascending the red giant
branch. The planet has an eccentric orbit ($e = 0.535 \pm 0.030$)
with a period of $52.50097\pm0.00021$\,d. After Kepler-56\,b,c and
Kepler-91\,b, Kepler-432\,b becomes the fourth known transiting
planet orbiting an evolved star. These planets have quite
different characteristics from those detected by the RV method,
and cover regions of parameter spaces that were considered to be
deserts until now, see Fig.\ \ref{fig:diagram}. They are also
important indicators of the formation processes and evolutionary
scenarios for planets around early-type stars.

\citet{mazeh2013} found no significant transit-time variations for
Kepler-432, and our RV data do not show any hint of a trend caused
by a longer-period companion. More RVs and a longer time-span are
necessary to constrain the possible presence of a third body that
might be responsible for the location and eccentricity of
Kepler-432\,b.

Since Kepler-432\,A is still evolving and expanding, this planetary
system is also very interesting from a dynamical point of view.
Currently, the planet reaches the minimum distance of $7.29 \pm
0.52\,R_\star$ at periastron, while at apastron is $24.08 \pm 0.85
\,R_\star$ away. However, at the end of the red giant branch, the
star will have a radius of $\sim 8\,R_{\sun}$ and, if we exclude a
possible orbital decay due to angular momentum transfer mechanism,
the distance of the planet from the star at periastron will be
$\sim$$3.8\,R_\star$. This means that the planet will not be
devoured by its parent star, but will instead accompany it towards
a more distant common fate.

\hspace{0.5cm} \linebreak The present letter was contemporaneously
submitted with that by Ortiz et al., who also confirm the
planetary nature of Kepler-432\,b.

\begin{acknowledgements}
Based on observations obtained with the CAHA 2.2 m Telescope and
the publicly available data obtained with the NASA space satellite
\emph{Kepler}. The Keck images used in this work were taken by
David Ciardi. SC\ thanks D.\ Gandolfi, M.\ I.\ Jones, L. Fossati,
and M.\ Ortiz for useful discussion.
\end{acknowledgements}

\bibliographystyle{aa}


\clearpage
\appendix
\section{Radial velocity measurements}

\begin{table}[h]
\centering %
\caption{BVS and RV measurements of Kepler-432.}
\begin{tabular}{cccc}
\hline %
\hline %
Date of observation  & BVS & RV    & err$_{\mathrm{RV}}$ \\
BJD-2450000          & m\,s$^{-1}$ & km\,s$^{-1}$ & km\,s$^{-1}$        \\
\hline %
6418.5145 &  32.43 & -35.394 & 0.061  \\
6418.5390 &  96.87 & -35.404 & 0.034  \\
6426.4901 &  46.53 & -35.948 & 0.070  \\
6426.5199 & 283.46 & -35.887 & 0.079  \\
6431.5465 & 153.41 & -35.893 & 0.052  \\
6432.6174 &  72.63 & -35.939 & 0.047  \\
6435.5761 &  81.90 & -35.771 & 0.035  \\
6435.6233 &  54.48 & -35.850 & 0.036  \\
6496.6430 &  73.72 & -35.915 & 0.057  \\
6501.6327 & 333.03 & -35.814 & 0.126  \\
6505.4502 & 160.12 & -35.738 & 0.044  \\
6505.4721 &  94.86 & -35.757 & 0.032  \\
6513.3674 &  43.42 & -35.434 & 0.063  \\
6514.5522 &  76.16 & -35.449 & 0.071  \\
6514.5742 & 114.34 & -35.676 & 0.076  \\
6515.5659 & 150.76 & -35.545 & 0.080  \\
6515.6022 & 249.71 & -35.617 & 0.063  \\
6524.5087 & 171.76 & -35.454 & 0.051  \\
6811.6033 & 213.03 & -35.924 & 0.053  \\
6811.6253 &  87.62 & -35.842 & 0.056  \\
6818.4792 & 179.78 & -35.869 & 0.057  \\
6818.5115 &  74.71 & -35.805 & 0.055  \\
6824.4574 & -26.04 & -35.620 & 0.066  \\
6824.4794 &  27.84 & -35.534 & 0.069  \\
6842.6148 & 222.54 & -35.849 & 0.070  \\
6842.6368 & 181.43 & -35.669 & 0.072  \\
6843.3949 & 142.63 & -35.795 & 0.057  \\
6843.4168 &  98.29 & -35.829 & 0.057  \\
\hline %
\end{tabular}
\label{ObsLog}%
\end{table}

\end{document}